# A Theoretical Analysis on Enthalpy of Vaporization: Temperature-Dependence and Singularity at the Critical State

Dehai Yu and Zheng Chen[*]

*BIC-ESAT, SKLTCS, CAPT, Department of Mechanics and Engineering Science, College of Engineering, Peking University, Beijing 100871, China*

**Abstract**

Accurate evaluation of enthalpy of vaporization (or latent heat of vaporization) and its variation with temperature is of great interest in practical applications, especially for combustion of liquid fuels. Currently, a theoretically rigorous formulation for latent heat is still not in place; and the existing fitting formulas for various fluids are, in nature, system-dependent. In this paper, the enthalpy of vaporization and its change with temperature are derived rigorously from first principles at the reference and critical states. A composite formula for enthalpy of vaporization is constituted from physical plausibility. The composite formula contains no fitting parameters; and it can accurately predict the enthalpy of vaporization from the reference temperature to the critical temperature.

## 1. Introduction

In most thermal engineering facilities, liquid fuels are used while chemical reactions in fact occur in gas phase [1, 2]. Therefore, vaporization of liquid fuel is crucial during its combustion [1, 3-5]. At constant temperature and pressure, equilibrium vaporization absorbs a certain amount of energy, which facilitates the liquid molecules escaping from the attractive potential induced by molecular interactions and is accordingly defined as the enthalpy of vaporization [6]. The vaporization rate depends strongly on the latent heat: the lower the latent heat, the higher the vaporization rate [1].

In the literature on multi-phase combustion [1, 5, 7, 8], the enthalpy of vaporization was widely assumed as a constant in theoretical treatment. However, this assumption may not hold rigorously during the combustion process with significant increase in temperature and pressure [1, 8-11]. For instance, ignition usually occurs when the system is initially at room temperature and atmospheric pressure with proper ignition energy supplied to the combustible reactants [3, 4, 8, 12]. As the combustion approaches the steady state, the ambient temperature increases significantly due to the exothermic reaction; and the pressure could also undergo considerable rise for combustion in a closed chamber [10, 13]. Hence, the assumption of constant latent enthalpy tends to be physically implausible.

---

[*] Corresponding author
  email address: cz@pku.edu.cn

Experiments have been extensively conducted to measure the latent heat of various fluids [14-22]. The enthalpy of vaporization was found to undergo increasingly reduction with temperature [19, 23]. Particularly, it vanishes at the critical state, where the phase distinction disappears. More interestingly, the gradient of latent heat with respect to temperature tends to be infinite at the critical state [3, 19, 23, 24]. The latent heat could be fitted from existing experimental data [19, 23]. However, these fitting formulas are system-dependent with many fitting coefficients, and they cannot reveal physical insights and lack generality. Hsieh et al. [4] numerically determined the latent heat by solving the Clausius-Clapeyron equation in terms of fugacity. The calculated latent heat agrees well with experimental data. However, there is no analytical expression derived for latent heat [2, 4].

To enhance physical understandings upon vaporization and to accurately calculate the vaporization rate of various fluids, a theoretical analysis on the latent heat is needed. In this study, we derived an analytical formula for latent heat which is applicable to various fluids. Moreover, this formula correctively interprets the latent heat from the reference temperature (room temperature in the present work), $T_r$, to the critical temperature, $T_c$. The choice of $T_r$ is determined by the availability of experimental data of latent heat. The theoretical formulation initiates from the principle of energy conservation at molecular level. The characteristic of vaporization changes noticeably from the reference state to the critical state [2, 4, 24, 25], implying that the formulation should be conducted separately at these two states. At the reference state, the configurational distribution of liquid molecules is characterized by the coordination number, $q$, i.e. the number of nearest-neighboring molecules surrounding the concerned molecule. Knowing the coordination number, the molecular heat capacity in liquid phase, $c_{v,l}$, can be appropriately evaluated. Subsequently, the latent heat at the reference state, $L_R$, is derived as a function of temperature $T$ and pressure $p$ (correlating to intermolecular distance, $r$). Near the critical state, the molar heat capacities, $C_v$, and density difference, $\rho_l - \rho_g$, are related to $T - T_c$ as power functions with specific exponents, which are called critical exponents. Accordingly, derivation of latent enthalpy near critical state, $L_C$, involves critical exponents, which ensures both $L_C = 0$ and $dL_C/dT = \infty$ at $T = T_c$. To interpret the latent heat in the temperature range from $T_r$ to $T_c$, a composite formula $L_I$ is constituted by properly weighting $L_R$ and $L_C$ in accordance with temperature variation. The composite formula can reduce to $L_R$ and $L_C$ at $T = T_r$ and $T = T_c$, respectively. Furthermore, good agreement with the experimental data is achieved in the concerned temperature range.

## 2. Formulation

### 2.1 Latent heat at molecule level

From statistical physics [6], the latent heat is the enthalpy difference between gas and liquid molecules during the phase transition, i.e.

$$l = h_g - h_l \tag{1}$$

where the subscripts $l$ and $g$ denote the liquid- and gas-phase, respectively. In thermodynamics, the enthalpy is related to the internal energy $\varepsilon$ through

$$h = \varepsilon + pv \tag{2}$$

where $v$ is the volume occupied by each molecule. In general, internal energy includes both individual molecular energy $\varepsilon_k$ and intermolecular potential energy $\varepsilon_p$, i.e. $\varepsilon = \varepsilon_k + \varepsilon_p$. The former is determined in terms of molecular heat capacity $c_v$,

$$\varepsilon_k(T) = \varepsilon_k(T_c) + \int_{T_c}^{T} c_v \, dT \tag{3}$$

It should be noted that the distinction between the liquid and gas phases disappears at the critical state, which leads to $\varepsilon_{g,k}(T_c) = \varepsilon_{l,k}(T_c)$.

The intermolecular potential energy pertains to the interacting forces among molecules, which is the mechanism making liquid molecules closely arranged [26-28]. It can be written as

$$\varepsilon_p = \frac{1}{2} q_i \Gamma(r_i), \qquad i = l, g \tag{4}$$

The factor $1/2$ is because every molecule should be treated equally and the interacting potential energy is shared by a pair of molecules. Substituting (2) - (4) into (1) yields

$$l = \int_{T_c}^{T} (c_{v,g} - c_{v,l}) \, dT + pv_g - pv_l + \frac{1}{2} q_g \Gamma(r_g) - \frac{1}{2} q_l \Gamma(r_l) \tag{5}$$

Equation (5) is the general form of molecular latent heat. Evaluation of $l$ requires (a) an appropriate potential function, (b) the constant-volume heat capacity per molecule, and (c) the molecular volume, for both liquid- and gas-phases. These quantities pertain to fundamental properties of the fluid, which implies the general validity of equation (5).

For simplicity, we assume that the fluid is non-polar. According to molecular thermodynamics, the intermolecular potential could be understood as interaction between induced dipoles [29]. The attractive potential can be rigorously derived as $\Gamma_{\text{att}} \sim -r^{-6}$. The repulsive force characterizes the incompressibility of molecules under normal conditions, which has negligible effect during vaporization due to its exceedingly short range of action [27, 29]. For physical plausibility and mathematical convenience, we adopt the well-known Lennard-Jones potential to interpret the molecular interaction,

$$\Gamma(r) = \varepsilon_0 \left[ \left(\frac{r_m}{r}\right)^{12} - 2\left(\frac{r_m}{r}\right)^6 \right] \tag{6}$$

where $r_m$ is the equilibrium separation distance at which the attractive and repulsive forces are in balance [30], and $\varepsilon_0$ is the depth of the potential well. Approximately $r_m$ is the intermolecular distance in liquid. However, such estimation is plausible only at moderate temperature since thermal expansion could be considerable at sufficiently high temperatures. For instance, at the critical state, it has $r_{l,c} = r_{g,c} = r_c$ with critical separation $r_c$ considerably larger than $r_m$. For $r \to \infty$, it yields $\Gamma(r \to \infty) \to 0$. Accordingly, we argue that $r_m/r_g \ll 1$, since the magnitude of interacting potential in gas phase is negligible comparing with $|\varepsilon_0|$.

It is straightforward to calculate $c_{v,g}$, which is entirely determined by the excited degrees of freedom of each molecule. For a polyatomic molecule comprising $n$ atoms, the degrees of freedom are distributed as 3 for translational motion, 3 (or 2 for linear molecule) for rigid rotation and the remaining $3n - 6$ (or $3n - 5$ for linear molecule) for internal vibration among atoms. According to the principle of equipartition [6, 27], we have

$$c_{v,g} = 3k + f_{v,g}(3n - 6)k, \qquad c'_{v,g} = \frac{5}{2}k + f'_{v,g}(3n - 5)k \tag{7}$$

where the prime indicates that the molecule has linear structure. The parameter $f_v$ represents the percentage of the excited internal vibrational degrees of freedom.

The physical scenario for liquid alters drastically, and it can be regarded as an intermediate state between gas and solid. The high fluidity and low resistance to shearing strain shows its gas-like properties; while the high density of liquid and the resulting multi-body interactions between molecules classifies its property as solid-like. Formal treatment of multi-body interaction is to introduce the potential of the whole system of $N$ parties, $U(\vec{r}_1, \cdots, \vec{r}_N)$, where $\vec{r}_i$ refers to the coordinate of the $i^{th}$ molecule. Such treatment is impracticable for a macroscopic system [27, 31]. It is recognized that intermolecular forces are short-ranged, and that only nearest neighboring molecules participate the interaction [26, 27, 29, 31]. For simplicity, we can assume that the interaction potential is pair-wise additive [29, 31], i.e. $U(\vec{r}_1, \cdots, \vec{r}_N) = \sum_{i<j} \Gamma_{ij}$, where $\Gamma_{ij}$ represents interacting potential between molecular $i$ and $j$. The multi-body interaction can be estimated, by virtue of coordination number, as the sum of binary interactions.

The coordination number in gas phase is unity, $q_g = 1$, since there are only bimolecular interactions. In liquid phase, $q_l$, can be determined from the radial distribution functions [26, 27, 29], $g(r)$, which involves the configuration partition function, $Z_N = \int \cdots \int e^{-U_N/kT} d\vec{r}_1 \cdots d\vec{r}_N$, and it becomes impractical in macroscopic level. Although $g(r)$ can be approximately determined from the Kirkwood integral equation [27, 29, 32], the mathematical solution becomes exceedingly complicated and an explicit analytical solution cannot be obtained. The rigorous calculation of $q$ is prohibited by the mathematical complexity of intermolecular potential. However, an estimation of $q$ can be performed. The disappearance of cohesion among molecules leads to even distribution of molecules over the whole range of separation. In analogy to the closest packing of neutral, rigid spheres, it suggests that $q_l = q_{HS} = 12$, which is qualitatively consistent with that in the literature [26, 27, 29].

Compared to gas phase, the interacting force among liquid molecules is much stronger. The liquid molecules can be considered as being constrained by $q$ "springs", in analogy to the bond among solid molecules [26, 27, 33]. Consequently, $c_{v,l}$, consists of two parts. The first part pertains to the degrees of freedom of individual molecule, which is analogous to gas phase. The second one pertains to intermolecular oscillations, which is solid-like. According to the principle of equipartition, we have

$$c_{v,l} = 3k + f_{v,l}(3n-6)k + \frac{q_l}{2}k, \qquad c'_{v,l} = \frac{5}{2}k + f'_{v,l}(3n-5)k + \frac{q_l}{2}k \qquad (8)$$

where the factor $1/2$ is introduced for the same reason as in equation (5). It is noted that the relations (7) and (8) hold at moderate temperatures. For temperature close to the critical state, $c_v$ needs to be reformulated.

## 2.2 Molecular latent heat at the reference state

At moderate temperature, the molecular volumes for both phases are well separated, i.e. $v_m/v_g \ll 1$, and thereby higher order terms could be neglected. Moreover, we assume that the internal vibration is decoupled from intermolecular oscillation, suggesting $f_{v,g} \approx f_{v,l}$. Substituting (6) – (8) into (5), the molecular latent heat becomes

$$l_R = \frac{1}{2}q_l\varepsilon_0 - \int_{T_c}^{T} \frac{q_l}{2} k \, dT + kT - f_p p v_m \tag{9}$$

where $f_p = 1 + \varepsilon_0/2kT$ can be determined from the van der Waals equation of state [29]. The term, $\varepsilon_0/2kT \sim O(1)$ for most encountered gas in combustion [1], interprets the correction of $v$ due to the attractive force among gas molecules. Straightforward evaluation of $\varepsilon_0$ requires the zero-point energy of the molecule [34], and thereby it is beyond the scope of the present study. Alternatively, we introduce $l_0$ at the reference state

$$l_0 = \frac{1}{2}q_l\varepsilon_0 - \int_{T_c}^{T_r} \frac{q_l}{2} k dT + kT_r - f_p p_r v_m \tag{10}$$

Subtracting (10) from (9) yields

$$L_R = L_0 - \int_{T_r}^{T} \frac{q_l}{2} R \, dT + R(T - T_r) + f_p(p_r - p)V_m \tag{11}$$

where the capital letter refers to quantities in molar scale, e.g. $L_0 = N_A l_0$ and $V_m = N_A v_m$, with $N_A$ the Avogadro's number and $R = kN_A$ the universal gas constant. For various of fluids, $L_0$ is available from experimental data [14-22]. Equation (11) indicates that the pressure effect is negligible at the reference state since $f_p \sim O(1)$ and $pV_m/RT \sim O(10^{-3})$.

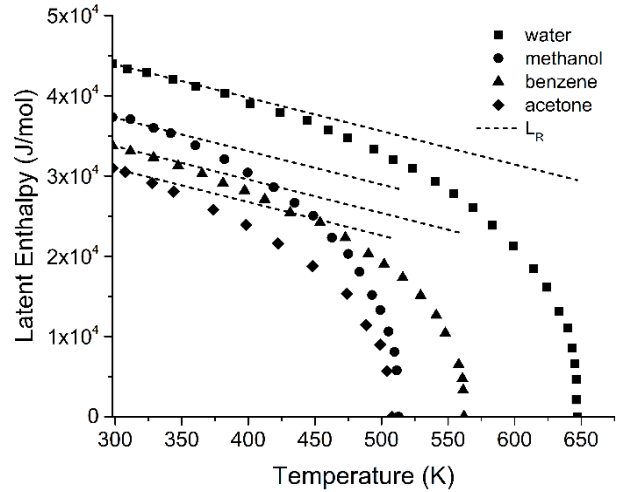

Figure 1. Comparison of theoretical prediction, $L_R$, with experimental data from [14-22].

At the reference state, $L_R = L_0$, whose value is specified from experimental data [14-22]. As shown in figure 1, the derivatives $dL_R/dT$ predicted by equation (11) agree well with experimental data. The decay of latent heat with temperature can be understood that with strong interaction, the liquid molecules can storage more energy for temperature increment $\Delta T$ due to the discrepancies in heat capacities at moderate temperature. Since the critical characteristic of the fluid has not been considered in $L_R$, Fig. 1 shows that $L_R$ deviates from experimental data significantly when the temperature increases toward the critical state.

**2.3 Molecular latent heat near the critical state.**

Correct interpretation to the latent heat near the critical state requires revision to the above formulation from two aspects. First, vanishing of phase distinction suggests equal molecular volumes, $v_{l,c} = v_{g,c}$, and identical molecular distribution, $q_{l,c} = q_{v,c}$. Accordingly, the discrepancy in the intermolecular potential energy disappears. Second, the density difference $\rho_l - \rho_g$ and the heat capacity $C_v$, can be interpreted as a power function of $T$,

$$\rho_l - \rho_g = f_\rho \rho_c \left(\frac{T - T_c}{T_c - T_r}\right)^\beta, \qquad C_v \sim f_{c,i} R \left(\frac{T - T_c}{T_c - T_r}\right)^{-\alpha} \qquad (12)$$

where $\rho_c$ (critical density) and $R$ are introduced for dimensional consideration [31, 35]. The parameters $\alpha$ and $\beta$ are the critical exponents, which are slightly system-dependent. The critical exponents can be approximately solved either from mean field theory [31, 35] or through molecular dynamics adopting the Ising model [25, 31]. More accurate calculation involves the renormalization group theory at the expense of exceedingly involved mathematics and physics [31]. Literature on the critical phenomena [6, 35] suggests that the values of $\alpha$ of $\beta$ for various fluids are approximately $1/8$ and $0.3$, respectively.

For a quantitative calculation of latent heat close to the critical state, $L_C$, a few parameters should be estimated. At moderate temperature, the factor $f_{c,v}$ deviates from $f_{c,l}$ by a quantity of $q_l R/2$. We assume that such discrepancy approximately remains up to the critical state, i.e. $f_{c,g} - f_{c,l} \approx -q_l R/2$. For different fluids, the critical density approximately satisfies $\rho_c/\rho_l \sim O(10^{-1})$, and the magnitude of $[(T - T_c)/(T_c - T_r)]^\beta$ lies in the interval $(0, 1)$. Subsequently, the magnitude of $f_\rho$ ranges from $O(10)$ to $O(10^2)$. Physical plausibility requires that $q_g < q_c < q_l$, and we approximately estimate $q_C$ as $q_c = \sqrt{q_g q_l} \approx 5$. The quantity $\varepsilon_0$ can be solved from equation (10): $\varepsilon_0 \approx 2l_0/q_l - k(T_c - T_r)$. Finally $L_C$ can be formulated as

$$L_C \approx \frac{q_l R}{2} \int_{T_c}^{T} \left(\frac{T_c - T}{T_c - T_r}\right)^{-\alpha} dT + \left[L_0 - \frac{q_l}{2} R(T_c - T_r)\right] \left(\frac{T_c - T}{T_c - T_r}\right)^{\beta} \tag{13}$$

The density factor $f_\rho$ is evaluated from the physical plausibility that $L_C/L_0 \sim O(1)$ when equation (13) is evaluated at the reference state.

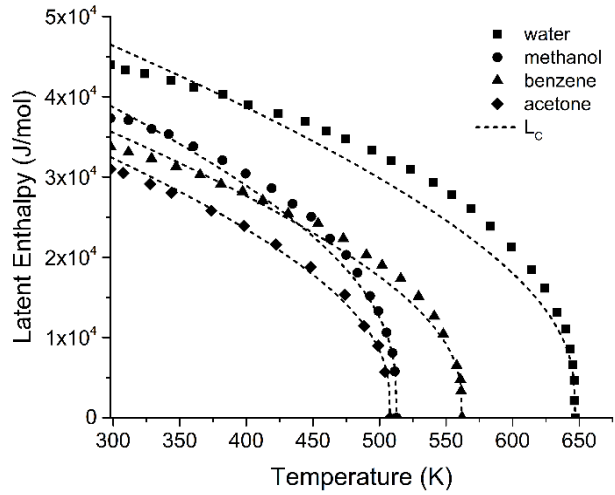

Figure 2. Comparison of theoretical prediction, $L_C$, with experimental data from [14-22]

At $T = T_c$, equation (13) gives $L_C = 0$. Moreover, the derivative $dL_C/dT$ is

$$\frac{dL_C}{dT} = \frac{q_l R}{2} \left(\frac{T_c - T}{T_c - T_r}\right)^{-\alpha} - \left[L_0 - \frac{q_l}{2} R(T_c - T_r)\right] \frac{\beta}{T_c - T_r} \left(\frac{T_c - T}{T_c - T_r}\right)^{\beta - 1} \tag{14}$$

which, according to $\alpha \approx 1/8$ and $\beta \approx 0.3$, becomes infinite at $T = T_c$. These behaviors of $L_C$ near the critical point are consistent with physical recognition and experimental data [3, 24, 25, 27, 35] as shown in figure 2. The singularity of $dL_C/dT$ at the critical state can be understood from two aspects. First, the statistical fluctuation among molecules [6, 31] becomes exceedingly significant, resulting in the divergence of heat capacity, i.e. $C_v \to \infty$ as $T \to T_c$. Second, the density difference shrinks at much more rapid rate than that for $T$ approaching to $T_c$, leading to a similar variation trend for discrepancy of intermolecular potential energy. However, as expected, $L_C$ cannot accurately predict latent heat at the reference state, where the critical expansions, equation (12), are invalid.

## 2.4 Integrated latent heat over the whole temperature range

It has indicated that $L_R$ and $L_C$ are valid only in limited temperature range. For practical interest, a composite formula, $L_I$, being valid from the reference state to the critical state, is needed. For physical plausibility, $L_I$ should satisfy the following conditions:

(a) $\|L_I - L_R\| \to 0$ at the reference state with $T \to T_r$;

(b) $\|L_I - L_C\| \to 0$ at the critical state with $T \to T_c$.

In analogy to expansions of (12) near the critical state, we introduce the normalized temperature $t = (T - T_r)/(T_c - T_r)$, and rewrite $L_R$ and $L_C$ as functions of $t$. The reference and critical states correspond to $t = 0$ and $t = 1$, respectively. It provides an approach to constitute the composite formula $L_I$ in terms of both $L_R$ and $L_C$. For mathematical convenience, we adopt the weighted-power-mean

$$L_I = L_R^{1-t} L_C^t \tag{15}$$

which satisfies both conditions (a) and (b). As $t \to 0$, the weight of $L_R$ becomes dominant over that of $L_C$ and thereby $L_I \to L_R$; and vice versa for $t \to 1$. Furthermore, the derivative $dL_I/dT$ is

$$\frac{dL_I}{dT} = (1-t)\frac{dL_R}{dT} L_R^{-t} L_C^t + t \frac{dL_C}{dT} L_R^{1-t} L_C^{t-1} \tag{16}$$

which reduces to $dL_R/dT$ for $t = 0$ at reference state, and to $dL_C/dT$ for $t = 1$ at the critical state. It is noted that $L_I$ has no fitting parameter. It is determined by $L_R$ and $L_C$, which can be evaluated respectively at the reference and critical states.

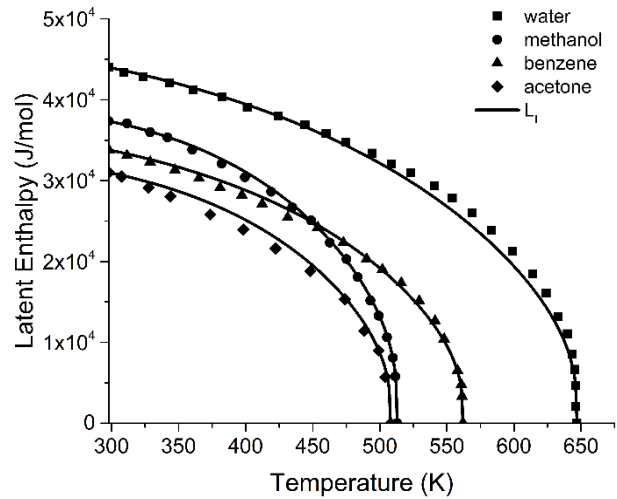

Figure 3. Comparison of the composite latent enthalpy formula (15) with experimental data from [14-22].

Figure 3 shows that the composite latent formula can qualitatively predict the latent heat of vaporization. The accuracy of $L_I$ might be slightly improved by either introducing fitting parameters or complicating its mathematical form. However, a rigorous solution must consider the physical mechanisms that bridge the reference state and the critical state. For instance, we suppose thermal expansion would be one of such bridging mechanisms, which can be interpreted as follows. The temperature increase leads to the increment in molecular separation distance and is accompanied by the elevation of intermolecular potential energy in liquid phase. Subsequently, latent heat tends to undergo accelerating reduction according to equation (5), which merits future study.

## 3. Concluding remarks

In this study, a theoretical analysis on the enthalpy of vaporization is presented. An explicit formula for latent heat near the reference state, $L_R$, is rigorously derived at the molecular level in terms of coordination number for liquid molecules. By virtue of critical exponents, the latent heat near the critical state, $L_C$, is appropriately interpreted as power functions of temperature deviation, $T - T_c$. For plausible interpretation of latent heat from the reference temperature to the critical temperature, a composite formula $L_I$ is constituted in terms of $L_R$ and $L_C$ with temperature dependent weighting exponents. The $L_I$ contains no fitting parameters. It can reduce to $L_R$ and $L_C$ respectively at the reference and critical states; and similar behavior applies to $dL_I/dT$. Furthermore, in comparison with experimental data, the composite formula accurately predicts the latent heat from the reference state to the critical state.

**Acknowledgement**

This work was supported by Beijing Innovation Center for Engineering Science and Advanced Technology.